\documentclass[a4paper,10pt,twoside]{cpc-hepnp}

\usepackage{url}
\usepackage{multicol}
\usepackage{graphicx}
\usepackage{booktabs}
\usepackage{amssymb,bm,mathrsfs,bbm,amscd}
\usepackage[tbtags]{amsmath}
\usepackage{lastpage}
\usepackage{lineno}

\usepackage[abs]{overpic}

\makeatletter
    
    \newcommand{\Rmnum}[1]{\expandafter\@slowromancap\romannumeral #1@}
\makeatother

\begin{document}


\fancyhead[c]{\small Submitted to Chinese Physics C}
\fancyfoot[C]{\small \thepage}

\footnotetext[0]{ Revised 22 May 2014}

\title{Study of diphoton decays of the lightest scalar Higgs boson in the Next-to-Minimal
Supersymmetric Standard Model
\thanks{Supported by National Natural
Science Foundation of China (No. 10721140381, No. 11061140514),
China Ministry of Science and Technology (No. 2013CB838700), China
Scholarship Council and partially by the France China Particle
Physics Laboratory}}

\author{
\quad FAN Jia-Wei$^{1,2,3}$
\quad TAO Jun-Quan$^{1}$
\quad SHEN Yu-Qiao$^{1,2}$ \\
\quad CHEN Guo-Ming$^{1}$ \quad CHEN He-Sheng$^{1}$
\quad S. Gascon-Shotkin$^{3}$
\quad M. Lethuillier$^{3}$ \\ 
\quad L. Sgandurra$^{3}$ \quad P. Soulet$^{3}$ }

\maketitle

\address{
$^1$ Institute of High Energy Physics, Chinese Academy of Sciences, Beijing 100049, China \\
\vspace*{4pt}
$^2$ University of Chinese Academy of Sciences, Beijing 100049, China \\
$^3$ Institut de Physique Nucl\'{e}aire de Lyon, Universit\'{e} de Lyon, Universit\'{e} Claude Bernard Lyon 1, CNRS-IN2P3,\\
     Villeurbanne 69622, France \\
}

\begin{abstract}
The CMS and ATLAS experiments at the LHC have announced the
discovery of a Higgs boson with mass at approximately
125~$GeV/c^{2}$ in the search for the Standard Model Higgs boson via
notably the  $\gamma\gamma$ and $ZZ$ to four leptons final states.
Considering the recent results on the Higgs boson searches from the
LHC, we study the lightest scalar Higgs boson $h_{1}$ in the
Next-to-Minimal Supersymmetric Standard Model by restricting the
next-to-lightest scalar Higgs boson $h_{2}$ to be the observed
125~$GeV/c^{2}$ state. We perform a scan over the relevant NMSSM
parameter space that is favoured by low fine-tuning considerations.
Moreover, we also take the experimental constraints from direct
searches, $B$-physics observables, relic density and anomalous
magnetic moment of the muon measurements as well as the theoretical
considerations into account in our specific scan. We find that the
signal rate in the two-photon final state for the NMSSM Higgs boson
$h_{1}$ with the mass range from approximately 80~$GeV/c^{2}$ to
122~$GeV/c^{2}$ can be enhanced by a factor up to 3.5, when the
Higgs boson $h_{2}$ is required to be compatible with the excess
from latest LHC results. This motivates the extension of the search
at the LHC for the Higgs boson $h_{1}$ in the diphoton final state
down to masses of 80~$GeV/c^{2}$ or lower, in particular with the
upcoming proton-proton collision data to be taken at center-of-mass
energies of 13-14 $TeV$.
\end{abstract}

\begin{keyword}
Supersymmetry, Next-to-Minimal Supersymmetric Standard Model,
Lightest scalar Higgs boson
\end{keyword}

\begin{pacs}
11.30.Pb, 14.80.Cp
\end{pacs}



\begin{multicols}{2}

\section{Introduction}
The Standard Model (SM) of particle physics has been very successful
in explaining high-energy experimental results. One of the remaining
questions is to find out what is the source of mass. The solution to
this question in the SM is given by the mechanism introduced by
Higgs, Englert and Brout\cite{labFANHiggsF,
labFANHiggsH,labFANHiggsG} which introduces an additional scalar
field whose quantum, the so-called Higgs boson, should be
experimentally observable. In July 2012, a Higgs boson-like particle
with mass at approximately 125~$GeV/c^{2}$ was announced to be
discovered by the two experiments, ATLAS and CMS, independently at
the LHC via notably the two most promising channels, $H \to
\gamma\gamma$ and $H \to ZZ^{*}$ channel with a four-lepton final
state\cite{labCMSHiggs,labATLASHiggs,labATLASNew,labCMSNew}.
Meanwhile, the Tevatron collaborations also announced their new
Higgs boson search results, based mainly on $VH$ associated
production with $H \to b \bar{b}$ decay
channel\cite{labTevatronHiggs}, which supported the LHC
$\sim$125~$GeV/c^{2}$ Higgs boson-like particle discovery results.
More data should be accumulated in order to test, with higher precision, the consistency between the data analysis results and the SM predictions on the signal strength. If a significant offset were to appear with a more precise measurement in the future, it could provide a window to new physics beyond the Standard Model (BSM).

Supersymmetry (often abbreviated
SUSY)\cite{labFANSUSYP,labFANSUSYF,labFANSUSYR,labFANSUSYM} is one
of the theoretical options for BSM physics. It introduces a new
symmetry between fermions and bosons. The most common SUSY framework
is the Minimal Supersymmetric Standard Model
(MSSM)\cite{labFANMSSM1,labFANMSSM2,labFANMSSM3} which keeps the
number of new fields and couplings to the minimum. In the MSSM, the
Higgs sector contains two Higgs doublets, which leads to a spectrum
including two CP-even, one CP-odd and two charged Higgs bosons. The
Lagrangian of the MSSM contains a supersymmetric mass term, the
$\mu$-term. This mass term is invariant under supersymmetry and
therefore it seems unrelated to the electroweak scale, although it
is phenomenologically required to be in this scale. This leads to
the well known "$\mu$ problem"\cite{labmuproblem,labmuproblemFAN} in
the MSSM. The simplest solution to this problem is the so-called
Next-to-Minimal Supersymmetric Standard Model (NMSSM)\cite{labNMSSM}
by introducing a new gauge singlet superfield which only couples to
the Higgs sector in a similar way as the Yukawa coupling and can
give rise to an effective $\mu$-term to solve the "$\mu$ problem".
Meanwhile, this new singlet adds additional degrees of freedom to
the NMSSM particle spectrum. In the CP conserving case, which is
assumed in this paper, the states in the Higgs sector can be
classified as three CP-even Higgs bosons $h_{i}$ ($i$ = 1,2,3), two
CP-odd Higgs bosons $a_{j}$ ($j$ = 1,2) and two charged Higgs bosons
$h^{\pm}$, for a total of seven observable states.

The extended parameter space of the NMSSM gives rise to a rich and
interesting phenomenology, in particular related to the two lightest
CP-even Higgs bosons $h_{i}$ ($i$ = 1,2). Inspired by the discovery
of the new particle with mass at approximately 125~$GeV/c^{2}$ from
the LHC and also the small LEP excess (approximately $2 \sigma$) at
about 98~$GeV/c^{2}$ in $e^{+}e^{-} \to Zh$ with $h \to b
\bar{b}$\cite{labLEPHiggs}, in this paper we study the lightest
CP-even Higgs boson $h_{1}$ in the mass range down to approximately
80~$GeV/c^{2}$, by assuming the next-to-lightest CP-even Higgs boson
$h_{2}$ as the new particle at m $\sim$125~$GeV/c^{2}$, along the lines of the studies of Belanger, Ellwanger, Gunion, Jiang, Kraml and Schwarz\cite{labnewUEll}. The third
CP-even Higgs boson is out of reach of current experiments due to
its low cross section 
in our scanned parameter ranges. 
To distinguish our study from many other NMSSM
studies\cite{labNSMMS1,labNSMMS2,labNSMMS3,labMarcin,labAndrea}, we
mainly focus on the regions of parameter space favoured by low
fine-tuning\cite{labNSMMS3,labCoup} considerations, with $\tan \beta$ chosen small, $\mu_{eff}$ chosen positive with minimal variations and low soft SUSY-breaking masses of the stop sector $M_{\widetilde{Q}_{3}}$ and $M_{\widetilde{t}_{R}}$. We choose an 
sbottom mass 
which is compatible with the SUSY search results at the
LHC\cite{labFANLHCSUSY1,labFANLHCSUSY2,labFANLHCSUSY3}. 
Furthermore, we perform our scan over the parameter space which can explain both dark
matter\cite{labPLANK}, ($g-2$)\cite{labDM} and some other
experimental constraints, described in section 3. For completely
testing the compatibility of our chosen region of parameter space
with the recent LHC results, we interface the package
$NMSSMTools$(version 4.1.0)\cite{labNMSSMTools,labNMSSMToolsa,labNMSSMToolsb,labNMSSMToolsc,
labNMSSMToolsd,labNMSSMToolse,labNMSSMToolsf,labNMSSMToolsg} with the newly
public packages $HiggsBounds\!-\!4$\cite{labHBound} and
$HiggsSignals\!-\!1$\cite{labHSignal}\footnote{
This work was begun with NMSSMTools version 3.2.4, which did not yet include the LHC H125 constraints.  We switched to version 4.1.0\cite{lanNMSSMLHCFAN}  to complete the work but for continuity chose to continue using the LHC H125 constraints from HiggsSignals\cite{labHSignal}. In NMSSMTools version 4.1.0 and later the LHC H125 constraints can be applied via explicitly requiring one of the NMSSM-predicted Higgs boson masses to lie within the interval $125.7 \pm 3 GeV/c^{2}$ and have signal rates compatible at the level of $2 \sigma$ with the combined signal strength ellipses of \cite{lanNMSSMLHCFAN}.}
Additionally, we show in section 4 that the $h_{2} \to
XX$ ($XX$ represents $\gamma\gamma$, $ZZ$, $WW$, $\tau\tau$, or
$bb$) signal strengths can be compatible with the current
experimental results, that signal strengths for an h1 with a mass
below 110 $GeV/c^{2}$ having higher values than currently predicted
by the Standard Model are possible, and that the current
sensitivities of the LHC experiments are such that the Higgs
boson h1 could be detected.

The structure of this paper is organized as follows. In section 2,
we briefly introduce the Higgs sector of the NMSSM. The details of
the parameter ranges we choose for the scan in the NMSSM parameter
space are described in section 3. Section 4 shows the results of our
numerical study including the Higgs boson $h_{2}$ signal strength in
each decay mode and the discussion on the lightest scalar Higgs
boson $h_{1}$. The summary and outlook are given in section 5.

\section{The NMSSM and Higgs boson signal strengths}

\subsection{Brief description of the NMSSM}

The general NMSSM includes two Higgs superfields $\hat{H_{u}}$,
$\hat{H_{d}}$ and one additional gauge singlet chiral superfield
$\hat{S}$. To start, we consider the NMSSM with a scale invariant
superpotential $W_{NMSSM}$ and the corresponding soft SUSY-breaking
masses and couplings $L_{soft}$, both of which are limited to the
R-parity and CP-conserving case. The superpotential $W_{NMSSM}$
depending on the Higgs superfields $\hat{H_{u}}$, $\hat{H_{d}}$ and
$\hat{S}$ is\cite{labNMSSM}

\begin{equation}
\begin{split}
\label{eq1}
W_{NMSSM} = h_{u}\hat{Q}\cdot\hat{H_{u}}\hat{U^{c}_{R}} + h_{d}\hat{H_{d}}\cdot\hat{Q}\hat{D^{c}_{R}} + h_{e}\hat{H_{d}}\cdot\hat{L}\hat{E^{c}_{R}} \\
            + \lambda\hat{S}\hat{H_{u}}\cdot\hat{H_{d}} +
            \frac{1}{3}\kappa\hat{S^{3}}.
\end{split}
\end{equation}

In the right-hand side of the above formula, the first three terms
are the Yukawa couplings of the quark and lepton superfields. The
fouth term replaces the $\mu$-term $\mu\hat{H_{u}}\hat{H_{d}}$ of
the MSSM superpotential. The last term, cubic in the singlet
superfield, is introduced to avoid the appearance of a Peccei-Quinn
axion, tightly constrained by cosmological
observation\cite{labNMSSM}. The corresponding soft SUSY-breaking
masses and couplings are given in the SLHA2\cite{labSLHA}
conventions by\cite{labNMSSM}

\begin{equation}
\begin{split}
\label{eq2}
-L_{soft} = m_{H_{u}}^{2}|H_{u}|^{2} + m_{H_{d}}^{2}|H_{d}|^{2} + m_{S}^{2}|S|^{2} + m_{Q}^{2}|Q|^{2} \\
           + m_{U}^{2}|U_{R}|^{2} + m_{D}^{2}|D_{R}|^{2} + m_{L}^{2}|L|^{2} + m_{E}^{2}|E_{R}|^{2} \\
           + h_{u}A_{u}Q \cdot H_{u}U_{R}^{c} - h_{d}A_{d}Q \cdot H_{d}D_{R}^{c} - h_{e}A_{e}L \cdot H_{d}E_{R}^{c} \\
           + \lambda A_{\lambda}H_{u}\cdot H_{d}S + \frac{1}{3}\kappa A_{\kappa}S^{3} +
           h.c..
\end{split}
\end{equation}

In Eq.\ref{eq1} and Eq.\ref{eq2}, clearly the non-zero vacuum
expectation value $s$ of the singlet $\hat{S}$ of the order of the
weak or SUSY-breaking scale gives rise to an effective $\mu$-term
with

\begin{equation}
\mu_{eff} = \lambda s .
\end{equation}
which solves the "$\mu$ problem" of the MSSM. Meanwhile, the three
SUSY-breaking mass-squared terms for $H_{u}$, $H_{d}$ and $S$
appearing in $L_{soft}$ can be expressed in terms of their VEVs
(Vacuum Expectation Value) through the three minimization conditions
of the scalar potential. Therefore, the Higgs sector of the NMSSM can be
described by the following six parameters

\begin{equation}
\lambda, \kappa, A_{\lambda}, A_{\kappa}, \tan{\beta}=\frac{\langle
H_{u}\rangle}{\langle H_{d}\rangle}, \mu_{eff} = \lambda \langle S
\rangle,
\end{equation}
in which each pair of brackets denotes the VEV of the respective
field inside them. In addition to these six parameters of the Higgs
sector, during the scan as described below we need to specify the
squark and slepton soft SUSY-breaking masses and the trilinear
couplings as well as the gaugino soft SUSY-breaking masses to
describe the model completely.

\subsection{Signal strength of Higgs boson}

As in the Standard Model (SM), the main Higgs boson production
processes include gluon-gluon fusion, vector boson fusion,
Higgs-strahlung and associated production with a vector boson or
$t\bar{t}$. The most dominant process is gluon-gluon fusion followed
by vector boson fusion 
and the other two associated production modes. In this paper, we will take all four production
processes into account.

We are interested in the Higgs boson signal strengths
$\mu_{XX}^{h_{i}}$ ($XX$ = $\gamma\gamma$, $ZZ$, $WW$, $bb$,
$\tau\tau$), which are the relative ratios of the cross section
times branching ratio ($R_{XX}^{h_{i}} = \sigma(pp \to h_{i}/pp \to h_{i}q\overline{q}/pp \to Vh_{i}/pp \to t\overline{t}h_{i}) \times
BR(h_{i} \to XX)$) to the SM predicted value: $\mu_{XX}^{h_{i}} =
R_{XX}^{h_{i}}/(R_{XX}^{h_{i}})_{SM}$.

In the NMSSM framework, the couplings of the Higgs bosons $h_{1}$, $h_{2}$ and $h_{3}$ depend on their decompositions into the CP-even weak
eigenstates $H_{d}$, $H_{u}$ and S, which are given
by\cite{labCoup}

\begin{equation}
\begin{split}
h_{1} = S_{1,d}H_{d} + S_{1,u}H_{u} + S_{1,s}S, \\
h_{2} = S_{2,d}H_{d} + S_{2,u}H_{u} + S_{2,s}S. \\
h_{3} = S_{3,d}H_{d} + S_{3,u}H_{u} + S_{3,s}S.
\end{split}
\end{equation}

\noindent where the coefficients $S_{i,u}, S_{i,d}$ quantify the amount of up-(down-) likeness and $S_{i,s}$ is a measure for the singlet component of a Higgs mass eigenstate.

Then the reduced tree-level couplings of $h_{i}$ ($i = 1, 2$) to $b$
quarks, $t$ quarks and electroweak gauge bosons $V$ relative to the
SM value are\cite{Ellwanger2010nf}

\begin{equation}
\begin{split}
\frac{g_{h_{i}bb}}{g_{h_{SM}bb}} = \frac{S_{i,d}}{\cos{\beta}},
\quad      \frac{g_{h_{i}tt}}{g_{h_{SM}tt}} = \frac{S_{i,u}}{\sin{\beta}},  \\
\frac{g_{h_{i}VV}}{g_{h_{SM}VV}} = \cos{\beta}S_{i,d} +
\sin{\beta}S_{i,u}.
\end{split}
\end{equation}


In the NMSSM, the coupling of $h_{1}$ to photons relative to that in the Standard Model is increased due to contributions from non-SM particles in the inducing loop diagrams. These can be from stop squarks but also to an even larger extent from charged Higgsinos where they are proportional to $\lambda$\cite{SchmidtHoberg, Choi}. In addition, the $h_{1}$ total width is smaller than that in the Standard Model due to a reduced coupling to b-quarks\cite{Ellwanger2010nf}. Both effects can serve to enhance the rate of the $h_{1}$ decay into two photons in some portions of parameter space.

\section{Scans with constrained parameters}

In the following, we will perform a specific scan in the NMSSM
parameter space 
which favours a Higgs boson h2 with a mass close to 125 $GeV/c^{2}$ and with
coupling strengths compatible with those measured at the LHC,
and the Higgs boson $h_{1}$ having mass restricted in the range down to
$\sim$80~$GeV/c^{2}$. The program package $NMSSMTools$ (version
4.1.0)\cite{labNMSSMTools,labNMSSMToolsa,labNMSSMToolsb,labNMSSMToolsc,
labNMSSMToolsd,labNMSSMToolse,labNMSSMToolsf,labNMSSMToolsg} is used to compute the SUSY particle and
NMSSM Higgs boson spectrum and branching ratios. $NMSSMTools$
contains four subpackages: $NMHDECAY$, $NMSDECAY$, $NMSPEC$ and
$NMGMSB$. The Fortran code $NMHDECAY$ provides the Higgs boson
masses, decay widths and branching ratios which will be used in this
paper. Furthermore, the $NMSSMTools$ package applies the constraints
from theory, 
constraints from direct Higgs boson searches at LEP\cite{labLEPHiggs}, Tevatron\cite{labTevatronHiggs} and LHC\cite{labCMS2013,labATLAS2013}, 
some bounds
from direct searches of SUSY particles in LHC
\cite{labFANLHCSUSY1,labFANLHCSUSY2,labFANLHCSUSY3}, relic density
$\Omega h^{2}$ \cite{labPLANK}, $B$-physics observables such as
BR($B\to X_{s}\gamma$), BR($B_{s}\to \mu^{+}\mu^{-}$),
BR($B_{\mu}\to \tau^{+}\nu_{\tau}$) and the mass mixings $\Delta
M_{s}$, $\Delta M_{d}$ \cite{labBPHY1,labBPHY2,labBPHY3,labBPHY4},
and anomalous magnetic moment of the muon ($g-2$)
constraints\cite{labDM}. All these constraints are used to perform
our scan. More details on the implementation of all these
constraints in the package can be checked from the webpage of the
$NMSSMTools$ program\cite{labNMSSMTools}.

After careful study, we have chosen to use the following parameter ranges motivated by the theoretical and experimental considerations detailed below. We realize that these may not be unique; more general ranges consistent with these considerations could possibly be more efficiently obtained via techniques such as Markov Chain Monte Carlo (MCMC) as described in \cite{labMCMC}.

\begin{enumerate}
\item To keep large doublet-singlet mixing in the Higgs sector,
we are more interested in large values of $\lambda$, $\kappa$ (but
small enough to avoid Landau pole below GUT scale), and to keep the fine-tuning as low as possible in a natural way, low values of
$\tan\beta$\cite{labCoup}. 
Considering the anomalous magnetic moment of the muon
($g-2$) constraint, we keep $\mu_{eff}$ positive with minimal
variations, in order to avoid fine-tuning\cite{labDM}. Hence the 4 parameters
are constrained in the following ranges

\begin{equation}
\begin{split}
0.6 < \lambda < 0.75, \quad 0.2 < \kappa < 0.3, \\
3 < \tan\beta < 4 , \quad 165 \textrm{$GeV/c^{2}$} < \mu_{eff} < 190
\textrm{$GeV/c^{2}$}.
\end{split}
\end{equation}

\item The soft SUSY-breaking trilinear couplings $A_{\lambda}$ and $A_{\kappa}$ are varied in the ranges\cite{labNMSSM}

\begin{equation}
\begin{split}
-100~GeV/c^{2} < A_{\kappa} < -50~GeV/c^{2}, \\
610~GeV/c^{2} < A_{\lambda} < 630~GeV/c^{2}.
\end{split}
\end{equation}

We remark that, constraining the parameters $A_{\lambda}$ and
$A_{\kappa}$ in these ranges favor $h_{1}$ with higher signal
strength as well as being in the mass range down to
$\sim$80~$GeV/c^{2}$.

\item In order to compare with the recent LHC search
bounds\cite{labFANLHCSUSY1,labFANLHCSUSY2,labFANLHCSUSY3}, we
conservatively set the left-handed soft SUSY-breaking masses of the
squark sector ($M_{\widetilde{Q}_{1,2}}$) and right-handed soft
SUSY-breaking sup masses ($M_{\widetilde{u}_{R}}$ and
$M_{\widetilde{c}_{R}}$) to 2500~$GeV/c^{2}$, both of which are in
the first two generations. We take low values of soft SUSY-breaking
masses of the slepton sector ($M_{\widetilde{L}_{1,2,3}}$,
$M_{\widetilde{e}_{R}}$, $M_{\widetilde{\mu}_{R}}$ and
$M_{\widetilde{\tau}_{R}}$) as 300~$GeV/c^{2}$ to follow the ($g-2$)
constraint\cite{labDM}. Furthermore, we set the right-handed soft
SUSY-breaking masses ($M_{\widetilde{D}_{R}}$) and the trilinear
couplings ($A_{D}$, $A_{E}$ and $A_{U}$) to 2500~$GeV/c^{2}$ and
1000~$GeV/c^{2}$ respectively. This results in light sbottom mass of
approximately $400~GeV/c^{2} < M_{\widetilde{b}_{1}} <
1000~GeV/c^{2}$ which is compatible with the recent LHC SUSY
results. Hence we have

\begin{equation}
\begin{split}
\label{eq10} 
M_{\widetilde{u}_{R}} = M_{\widetilde{c}_{R}} = M_{\widetilde{Q}_{1,2}} = 2500~GeV/c^2, \\
M_{\widetilde{L}_{1,2}} = M_{\widetilde{e}_{R}} = M_{\widetilde{\mu}_{R}} = 300~GeV/c^2, \\
M_{\widetilde{L}_{3}} = M_{\widetilde{\tau}_{R}} = 300~GeV/c^2, \\
M_{\widetilde{D}_{R}} = 2500~GeV/c^2 \quad (D = d,s,b), \\
A_{D} = A_{E} = 1000~GeV/c^2 \quad 
\\
A_{U} = 1000~GeV/c^2 \quad 
\\
\end{split}
\end{equation}

\item The Higgs sector is strongly influenced by the stop sector via radiative corrections\cite{labFANRA},
and also for fine-tuning reasons we further need to specify the soft
SUSY-breaking masses of the stop sector\cite{labCoup}. We modify the $NMSSMTools$
code in order to constrain them to be rather low. After studying the
properties of these parameters, we vary them simultaneously within

\begin{equation}
\label{eq11} 550~GeV/c^{2} < M_{\widetilde{Q}_{3}} =
M_{\widetilde{t}_{R}} < 700~GeV/c^2.
\end{equation}

(Eqs.\ref{eq10} and \ref{eq11} presuppose a SUSY scale.)

\item Concerning the relic density constraint\cite{labPLANK}, the remaining gaugino soft SUSY-breaking masses
are set to be within

\begin{equation}
\begin{split}
100~GeV/c^2 < M_{1} < 150~GeV/c^2, \\
180~GeV/c^2 < M_{2} < 300~GeV/c^2, \\
M_{3} = 1000~GeV/c^2.
\end{split}
\end{equation}

\end{enumerate}

Then we perform our scan after the application of the constraints on
the parameters as described above.

\section{Numerical study}

In section 2, we introduced the production processes and the signal
strengths of the NMSSM Higgs bosons. In this section, we demonstrate
that the constraints on the parameters as described in the above
section can produce a next-to-lightest NMSSM scalar Higgs boson
$h_{2}$ compatible with the observed state at the LHC with mass at
approximately 125~$GeV/c^{2}$. We concentrate our study on the
lightest NMSSM scalar Higgs boson $h_{1}$. Considering the relic
density $\Omega h^{2}$, we will focus on two cases, $\Omega h^{2} <
0.1102$ (named case $\Rmnum{1}$) and $0.1102 < \Omega h^{2} <
0.1272$ (the "WMAP" window\cite{labFANWMAP}, named case
$\Rmnum{2}$). In all plots below, points for case $\Rmnum{1}$ are
represented by blue squares and case $\Rmnum{2}$ by red triangles.

\subsection{Mass distributions of the NMSSM Higgs bosons }

Based on the constrained parameters, firstly we show the mass
distributions of the two lightest NMSSM scalar Higgs bosons $h_{1}$
and $h_{2}$ in Fig.~\ref{fig1}. As can be seen, most of the
parameter points cluster around mass values centered around 125
$GeV/c^2$ for $M_{h_{2}}$ in case $\Rmnum{1}$ and case $\Rmnum{2}$.
We conclude that the parameter ranges are correctly chosen to give a mass of the Higgs
boson $h_{2}$ close to 125 $GeV/c^2$.
Considering the lightest NMSSM scalar Higgs boson $h_{1}$,
it is clear that its mass can lie in a wide range,
approximately from 60 (70) $GeV/c^{2}$ to 122 $GeV/c^{2}$ for case $\Rmnum{1}$ (case $\Rmnum{2}$). We point out that 
the excluded region below 114.7 $GeV/c^2$ at LEP\cite{labLEPHiggs} could still be allowed in the context of the NMSSM for points in the parameter phase space where the production rate of $ee \to Z^{\ast} \to Zh_{1}$ with $h_{1}$ decaying
into bb or $\tau\tau$ (the channels searched for at LEP) are reduced or suppressed with respect to  the  SM.

\begin{center}
\includegraphics[width=8cm]{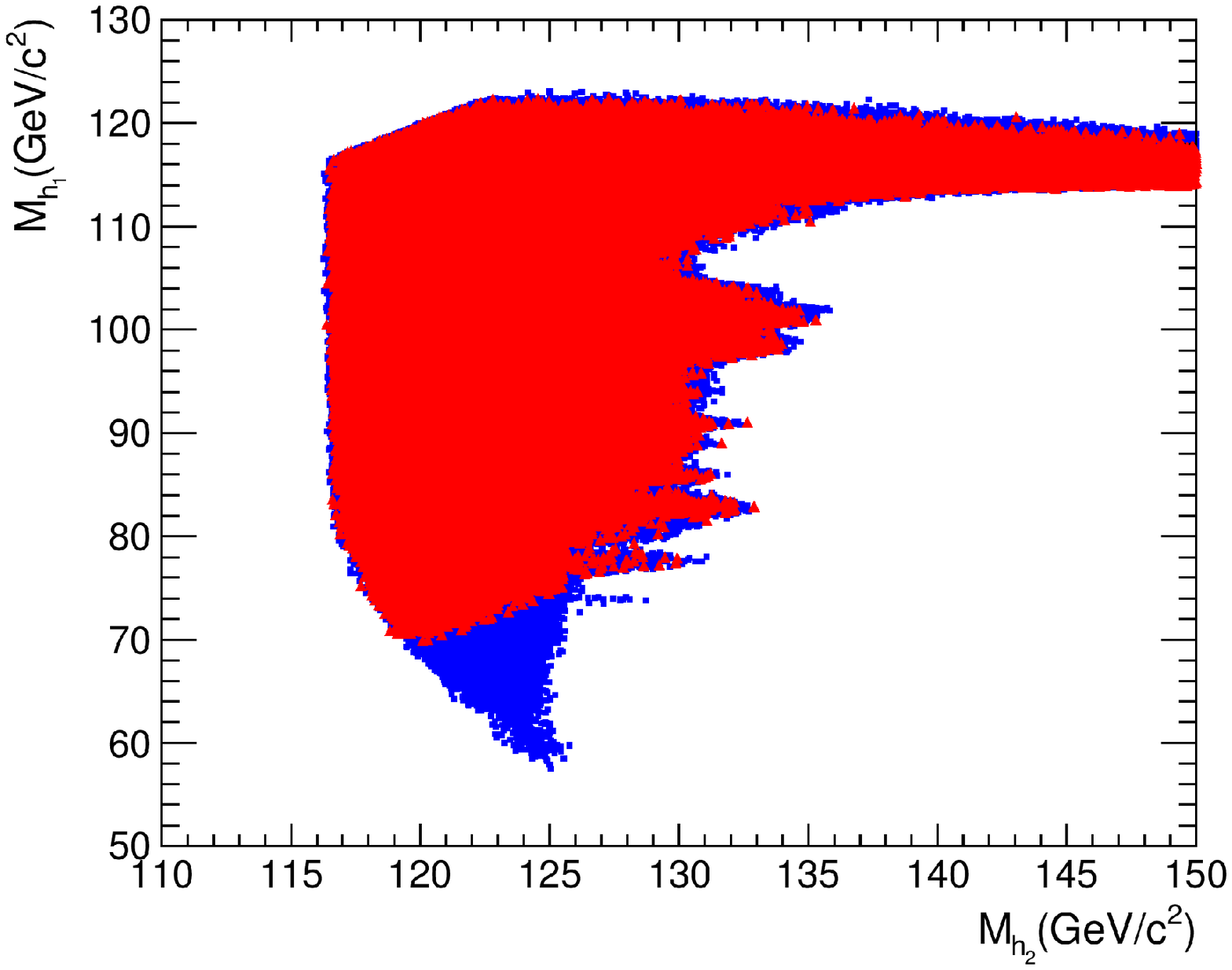}
\figcaption{\label{fig1} The NMSSM Higgs boson mass spectrum in the $M_{h_{1}}$ vs. $M_{h_{2}}$ plane. Points for case $\Rmnum{1}$ are represented by blue squares and case
$\Rmnum{2}$ by red triangles.}
\end{center}

\subsection{Signal strengths of the NMSSM Higgs boson $h_{2}$ }

In the NMSSM framework, not all the $\mu_{XX}^{h_{i}}$ (i=1, 2) are
independent, for example, $\mu_{ZZ}^{h_{i}} = \mu_{WW}^{h_{i}}$.
Only the reduced couplings are calculated by NMSSMTools, we use the
absolute values in the Standard Model\cite{labFANBR} to calculate
the total signal strength including four production modes mentioned
in section 2.2. We also check that the differences of weights of the
production mode are quite negligible between 7 and 8 $TeV$ with
repect to experimental uncertainties. In order to further test
whether a given point in our scanned parameter space is allowed or
excluded by the recent LEP, Tevatron and LHC results at $95\%$
confidence level (CL), two new public tools are utilised.

We use the public tool $HiggsBounds\!-\!4$\cite{labHBound} to
further compare Higgs sector predictions with existing exclusion
limits of various search channels. The SLHA format files calculated
by $NMSSMTools$ are used as the inputs for $HiggsBounds\!-\!4$. The
main algorithm of $HiggsBounds$ can be described in two steps. In
the first step, $HiggsBounds$ uses the expected experimental limits
from LEP, Tevatron and the LHC
\cite{labLEPHiggs,labTevatronHiggs,labCMS2013,labATLAS2013} to
determine which decay channel has the highest statistical
sensitivity. In the second step, only for this particular channel
the theory prediction is compared to the observed experimental
limits in order to conclude whether this parameter point is allowed
or excluded at $95\%$ CL.

Then, compatibility with the measured mass and rates of the observed
new state having a mass of $\sim$125~$GeV/c^{2}$ is imposed, using
the public code $HiggsSignals\!-\!1$\cite{labHSignal}.
$HiggsSignals\!-\!1$ takes the predictions of an arbitrary model
(here the NMSSM) as an input, providing a quantitative answer to the
statistical question of how compatible the model predictions are
with the Higgs boson search experimental results, especially signal
strengths and the measured mass, by evaluating a $\chi^{2}$
calculation. The main results from $HiggsSignals\!-\!1$, which are
used to further constrain our parameter space, are reported in the
form of a $\chi^{2}$ value and the associated $p$-value. We consider
that the given parameter point is compatible with the experimental
constraints only if the $p$-value given by $HiggsSignals\!-\!1$ is
greater than 0.05. By using these two programs in parallel, we
obtain the most complete test for the scanned NMSSM parameter space.

The allowed values for $\mu_{XX}^{h_{2}}$ from the scan over the
NMSSM parameter space are shown in Fig.~\ref{fig2}, where all the
constraints described in section 3 have been applied. The results
are shown before and after applying the additional constraints from
the above two programs. We first show  $\mu_{\gamma\gamma}^{h_{2}}$
plotted versus $\mu_{XX}^{h_{2}}$ ($XX$ = $ZZ$, $WW$, $bb$,
$\tau\tau$). The points including error bars represent the latest
LHC public results for the best fit values of the signal strengths
$\mu_{XX}^{h_{2}}$ with uncertainties in the different final states,
reported by the CMS and ATLAS
collaborations\cite{labCMS2013,labATLAS2013}. The values and errors
are listed in Table \ref{tab1} in the
Appendix\cite{labCMS2013,labATLAS2013,labATLASNew,labCMSNew}. It is
clearly visible that the parameter points compatible with both
$HiggsBounds\!-\!4$ and $HiggsSignals\!-\!1$ provide theory
predictions which are consistent with the experimental results. In
Fig.~\ref{fig2}, by taking the di-photon final state as an example,
we also show the signal strength $\mu_{\gamma\gamma}^{h_{2}}$
plotted against its mass in Fig.~\ref{fig2} (d). From the right-hand
plot, the $\mu_{\gamma\gamma}^{h_{2}}$ values cover the range 0.5 to
1.8 while the mass is in the range 120~$GeV/c^{2}$ to
132~$GeV/c^{2}$, both of which are consistent with the new observed
state within errors. It is clearly visible that the NMSSM can
produce rates compatible with both the CMS and ATLAS results for
both relic density cases. The plots show that the relatively sizable
enhancements with respect to the SM rates for the $\gamma\gamma$ and
$ZZ$ final states reported by ATLAS are possible in the vicinity of
125~$GeV/c^{2}$ in the NMSSM framework and also possible for the
relatively suppressed rates reported by CMS.

\begin{center}
\begin{minipage}[t]{0.45\textwidth}
\centering
\begin{overpic}[scale=0.4]{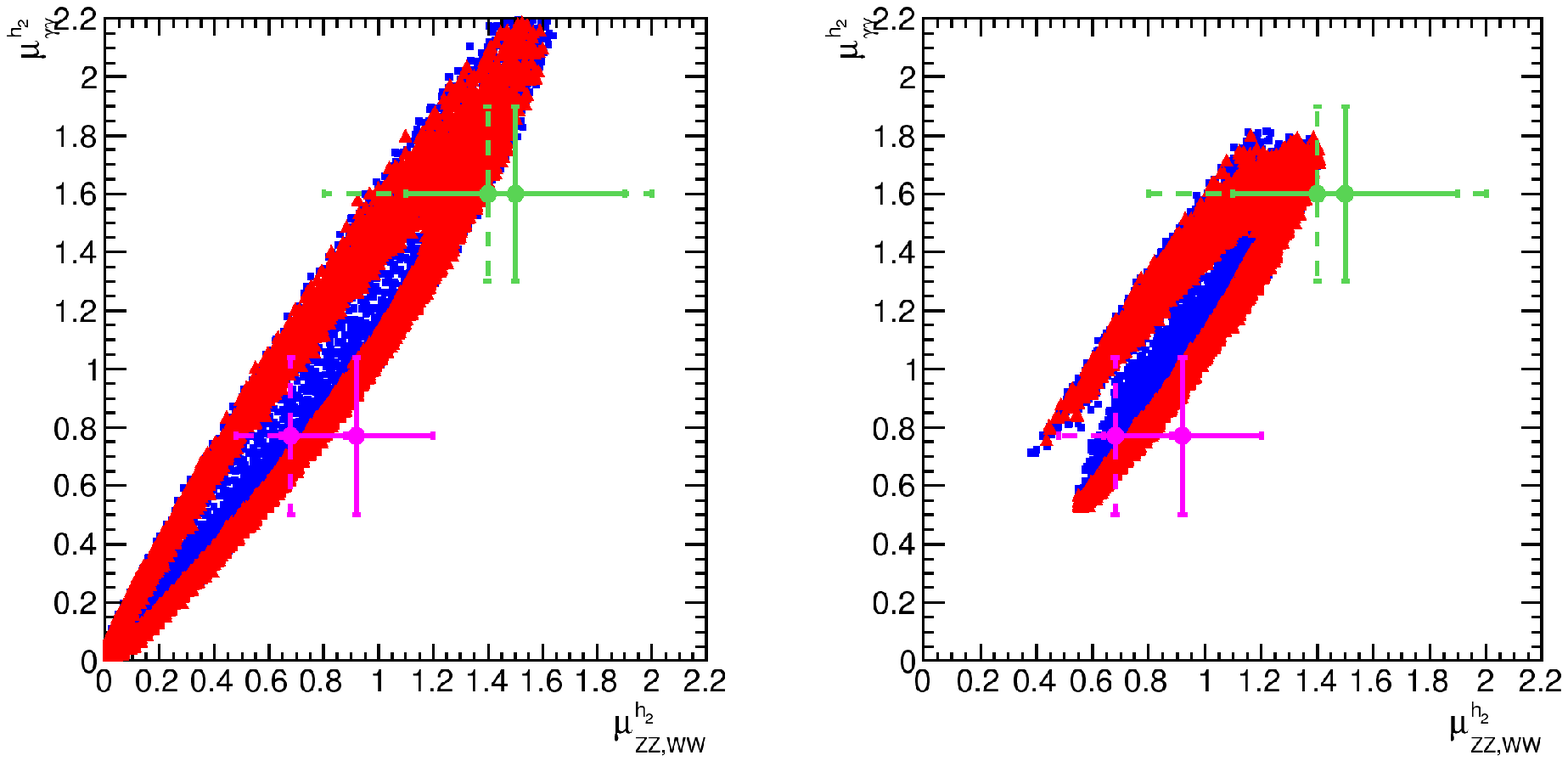}
\put(110,0){$(a)$}
\end{overpic}
\hfill
\end{minipage}
\begin{minipage}[t]{0.45\textwidth}
\centering
\begin{overpic}[scale=0.4]{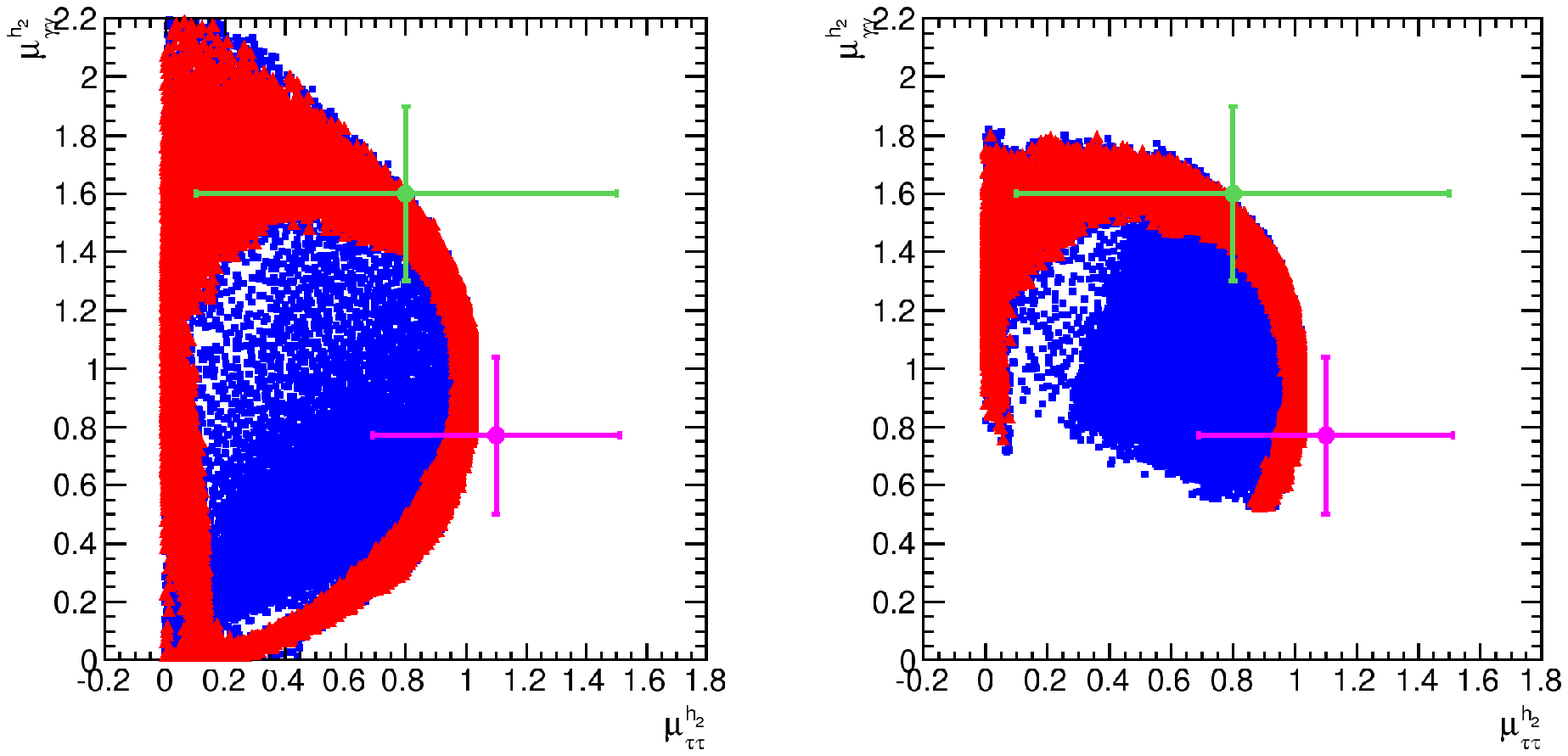}
\put(110,0){$(b)$}
\end{overpic}
\end{minipage}
\begin{minipage}[t]{0.45\textwidth}
\centering
\begin{overpic}[scale=0.4]{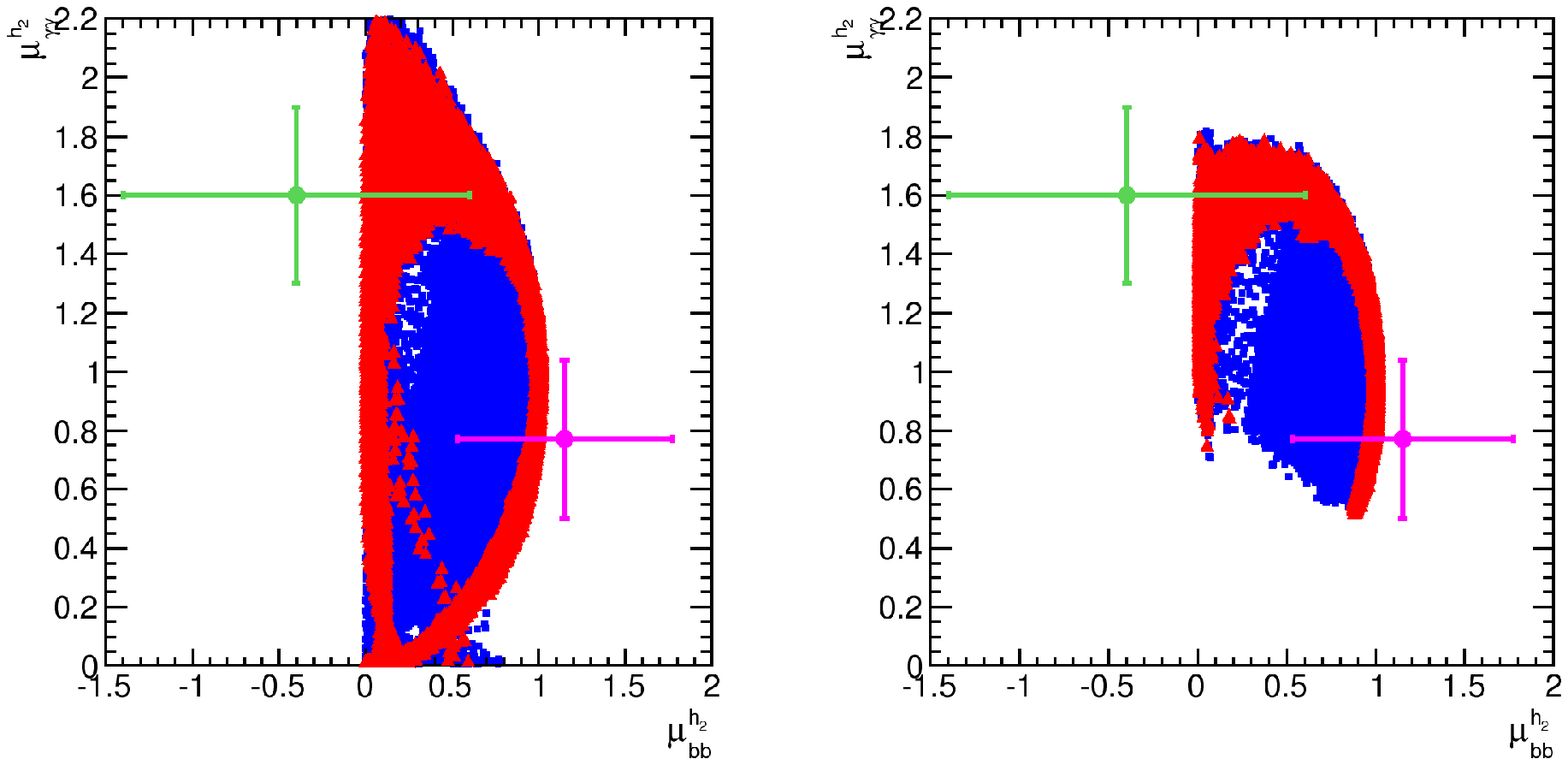}
\put(110,0){$(c)$}
\end{overpic}
\end{minipage}
\begin{minipage}[t]{0.45\textwidth}
\centering
\begin{overpic}[scale=0.4]{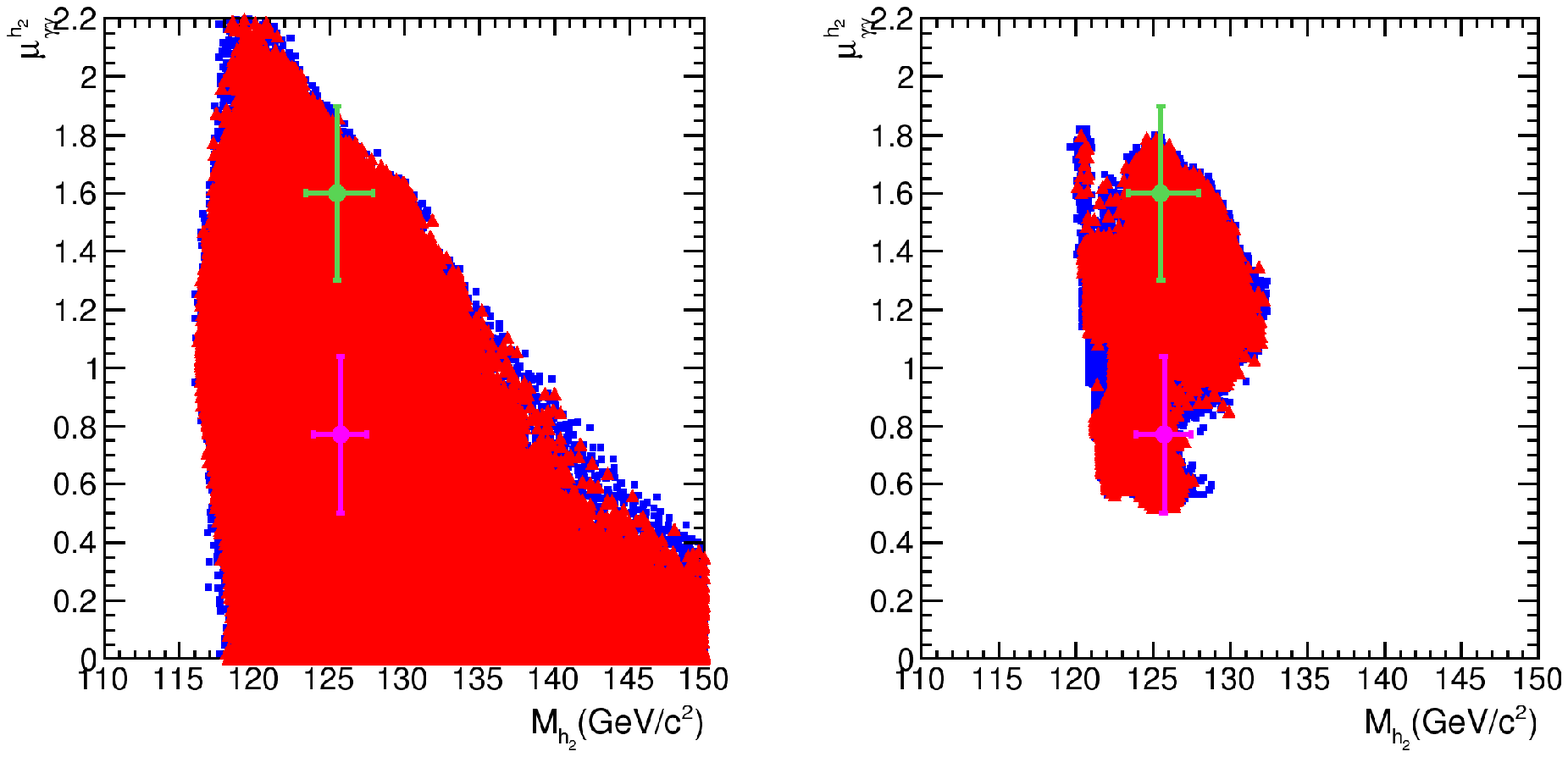}
\put(110,0){$(d)$}
\end{overpic}
\end{minipage}
\figcaption{ The signal strength of the NMSSM Higgs boson $h_{2}$ in the $\gamma\gamma$ channel versus its signal
strengths in the $ZZ$($WW$) channel in (a), the $bb$ channel in (b), the $\tau\tau$ channel in (c) and
its mass in (d). The results
shown in the left-hand (right-hand) plots are obtained before
(after) applying the constraints from $HiggsBounds\!-\!4$ and $HiggsSignals\!-\!1$. The
magenta solid point for the mean value and line for the
uncertainties represent the CMS results while those in green represent the
ATLAS results. Points for case $\Rmnum{1}$ are represented by blue
squares and case $\Rmnum{2}$ by red triangles. In (a), the solid
point and line represent the results from $ZZ$ final state while the
dashed line corresponds to $WW$ final state.} {\label{fig2}}
\end{center}

\subsection{Branching ratio and signal strength of the NMSSM Higgs boson $h_{1}$}

We will now focus our discussion on the lightest NMSSM scalar Higgs
boson $h_{1}$ by looking at the di-photon final state, and will
further restrict ourselves to the case of CMS results only.

Over and above the constraints mentioned in previous sections, we now demand in addition that the mass of the NMSSM Higgs boson $h_{2}$ be explicitly compatible with that most recently measured by CMS for the newly-discovered boson, and that the $h_{2}$ signal strength in the diphoton channel be compatible at a more stringent level with that measured by CMS.
In the most recent CMS results \cite{labCMSNew}, the SM-like Higgs boson mass has been measured to
be $125.7\pm0.3(stat.)\pm0.3(syst.)~GeV/c^2$. Assuming $3\sigma$
error, where $\sigma$ is taken as the linear sum of the above
statistical and systematic uncertainties, the mass of $h_{2}$ is
constrained within the range

\begin{equation}
      123.9~GeV/c^2 < M_{h_{2}} < 127.5~GeV/c^2.
\end{equation}

We also demand that the signal strength $\mu_{\gamma\gamma}^{h_{2}}$
should be within $1\sigma$ (taking as $\sigma$ the uncertainty shown
in the Appendix) of the CMS measured value:

\begin{equation}
      0.5  \lesssim \mu^{h_{2}}_{\gamma\gamma} \lesssim 1.04.
\end{equation}

Based on these additional constraints, Fig.~\ref{fig3} shows the
allowed values for the branching ratio of the $h_{1} \to
\gamma\gamma$ decay mode in the NMSSM. The cyan solid line shows the
quantity including error bands evaluated in the SM for the same
mass\cite{labFANBR}. From the plots, most of the points show that an
enhanced branching ratio relative to that in the SM is possible for
both relic density cases. The theory explanation for this enhanced
two-photons branching ratio has already been introduced in Section
2.2.

\begin{center}
\includegraphics[width=8cm]{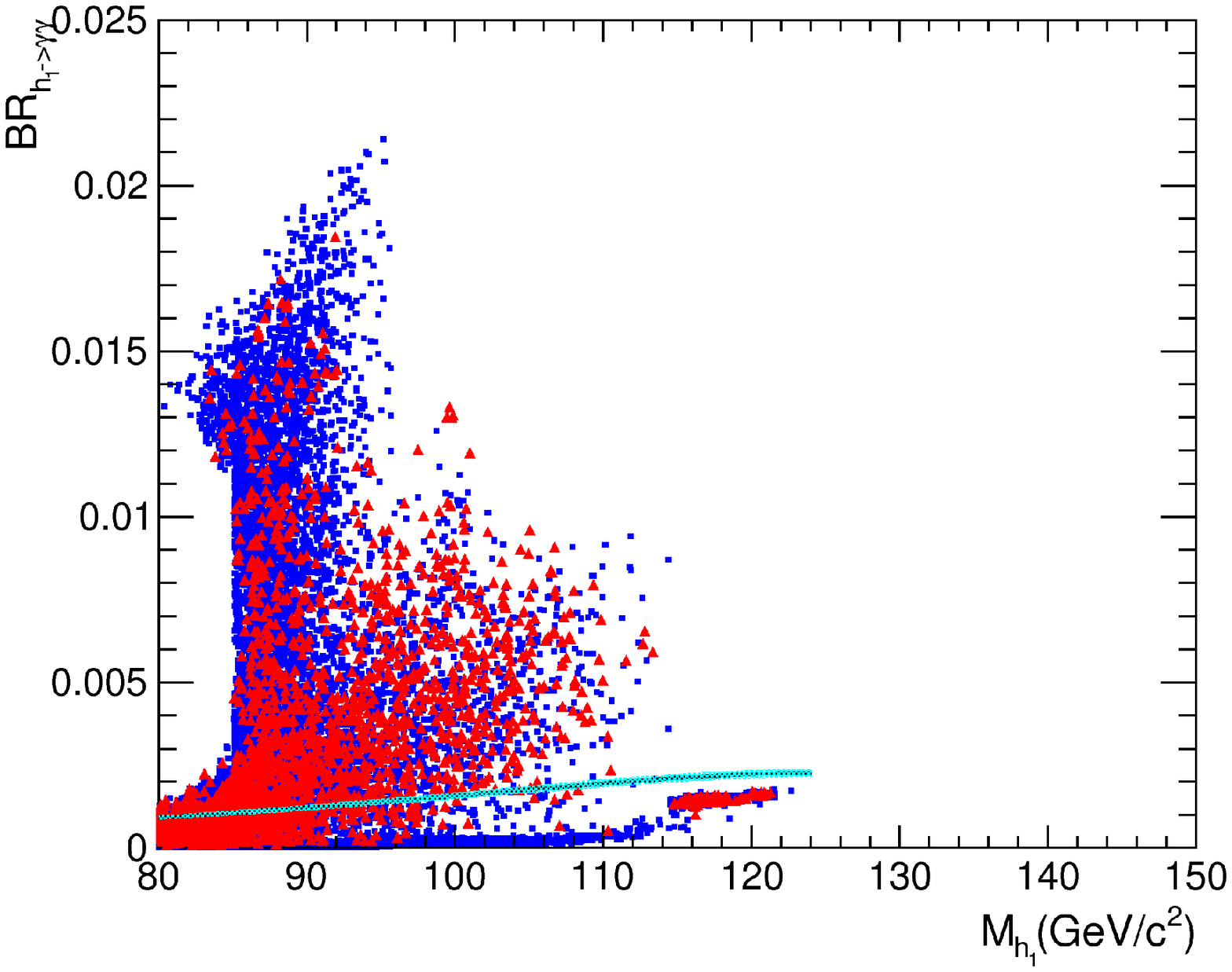}
\figcaption{ Results from the NMSSM parameter scan for the branching
ratio 
of $h_{1} \to \gamma\gamma$. The cyan solid line
represents the corresponding SM value for the same mass. Points for
case $\Rmnum{1}$ are represented by blue squares and case
$\Rmnum{2}$ by red triangles.}{\label{fig3}}
\end{center}

In Fig.~\ref{fig4}, we display the possible signal strengths
$\mu^{h_{1}}_{\gamma\gamma}$ plotted against the Higgs boson $h_{1}$
mass. As seen from Fig.~\ref{fig4}, the remaining points selected
after application of all the conditions discussed in Sections 3 and
4.2 indicate the possibility of the $h_{1}$ mass lying in the range
between 80~$GeV/c^{2}$ to 122~$GeV/c^{2}$ for both relic density
cases. Turning to the signal strength $\mu^{h_{1}}_{\gamma\gamma}$,
the figure shows that a sizable enhancement over the SM rate is
possible for the Higgs boson $h_{1}$ for both relic density cases,
reaching values as high as $3.5$, corresponding to an $h_{1}$ mass
of $\sim90$ $GeV/c^{2}$. We note that, for the mass range between
100~$GeV/c^{2}$ and 110~$GeV/c^{2}$, the allowed signal strengths
$\mu^{h_{1}}_{\gamma\gamma}$ are lower, falling to $\sim$0.9.

In order to compare with our signal strength
$\mu^{h_{1}}_{\gamma\gamma}$, we also superpose the official CMS
public exclusion limit plot in Fig.~\ref{fig4}. The yellow and green
regions correspond to the uncertainties at $95\%$ and $68\%$
confidence interval respectively and the cyan solid line corresponds
to the SM value. It is clearly seen that the NMSSM points above the
solid black line (representing the CMS observed exclusion limit) are
almost excluded by the CMS result in the mass range 110~$GeV/c^{2}$
to 122~$GeV/c^{2}$. We note that there is a small interesting region
that is favoured by a cluster of parameter points in the
neighborhood of 120~$GeV/c^{2}$. Especially, the points below
120~$GeV/c^{2}$ are already excluded by comparing with the solid
black line. The remaining points lying between 120 and
122~$GeV/c^{2}$ could constitute a case of a so-called "degenerate
Higgs boson"\cite{labDoubleHiggs} which is outside the scope of our
discussion in this paper. But it would  be advisable to test this
interesting scenario with the increased quantity of data and
improved resolution in the future 13 and 14-$TeV$ collisions at the
LHC. To date, the present experimental results from the LHC do not
cover the lower Higgs boson mass range between 80-110~$GeV/c^{2}$ in
the $H \to \gamma\gamma$ decay channel. 
In order to be able to make a conclusion for the NMSSM points in this mass range, which show the potential for sizeably enhanced signal strengths in the diphoton decay channel with respect to those predicted in the SM,  a detailed analysis is needed taking into account especially the $Z \to ee$  background faking the diphoton signals.
If the limit curve were to be extrapolated down to a mass of $\sim 80~GeV/c^{2}$
and the measurement on the signal strength improved in the future
experimental analysis, most of the NMSSM parameter space shown in
Fig. 4 could be probed.

\begin{center}
\includegraphics[width=6cm]{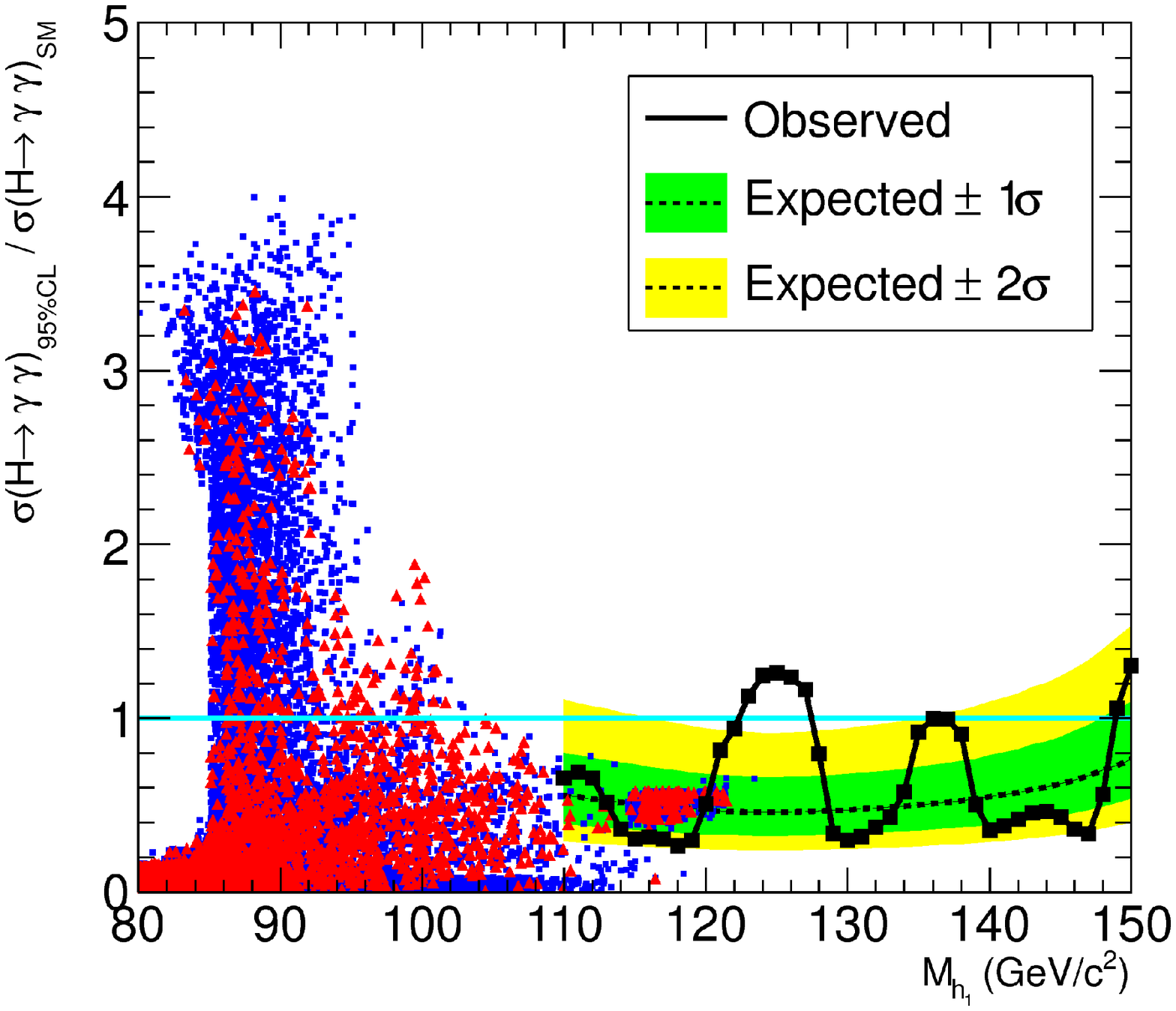}
\figcaption{ Expected and observed exclusion limits on the signal
strength from CMS\cite{labCMS2013} compared with the possible signal strengths of the process
$pp \to h_{1} \to \gamma\gamma$ from the NMSSM parameter scan. Points for case $\Rmnum{1}$ are
represented by blue squares and case $\Rmnum{2}$ by red triangles.The
solid black line together with the black squares corresponds to the ratio of the CMS observed
cross sections with respect to the SM predictions, and the dashed
line is the expected ratio. }
 {\label{fig4}}
\end{center}

\section{Summary and outlook}
In this paper, we have performed a scan in the NMSSM, focusing on
the regions of parameter space favoured by low fine-tuning
considerations. We have studied the lightest scalar Higgs boson
$h_{1}$ including the mass and the relative signal strength to the
SM prediction, especially for Higgs bosons decaying into the
di-photon final state, by assuming that the second-lightest scalar
Higgs boson $h_{2}$ corresponds to the observed
$\sim$125~$GeV/c^{2}$ state at the LHC. We find that a significant
excess of the signal strength relative to that of the Standard Model
in $pp \to h_{1} \to \gamma\gamma$ up to a factor $\sim$3.5 is
possible in the NMSSM, especially for the mass range below the LEP
bound of 114.7 $GeV/c^{2}$. We recommend that experiments extend the
exclusion limit to this low-mass region in order to investigate the
possibilities of the NMSSM in more detail.

With future LHC data, the best fit values of the signal strengths in
each channel may evolve and the uncertainties improve, which may
result in changes in the experimental results and reduced error bars
in our plots. Additionally, the allowed regions in NMSSM parameter
space for the interesting Higgs boson $h_{1}$ may also change. With
the upcoming 13 and 14-$TeV$ collisions at LHC, the signal for the
low-mass NMSSM Higgs boson $h_{1}$ could still well be detected by
the experiments due to the higher collision energy and integrated
luminosity.

\vspace{3mm}
\emph{The authors would like to thank Giacomo Cacciapaglia, Aldo Deandrea, Guillaume Drieu La Rochelle, Ulrich Ellwanger, Jean-Baptiste Flament, Jack Gunion, Cyril Hugonie, Yun Jiang and Sabine Kraml for helpful discussions.}

\end{multicols}


\vspace{-1mm} \centerline{\rule{90mm}{0.1pt}} \vspace{2mm}

\begin{multicols}{2}

\end{multicols}

\vspace{-1mm} \centerline{\rule{90mm}{0.1pt}} \vspace{2mm}

\section*{Appendix: Best fit values of the signal strength}


\vspace{3mm}

\begin{center}
\tabcaption{ \label{tab1} Best fit values ($\mu$) of the signal
strength reported by CMS and ATLAS
Collaborations\cite{labCMS2013,labATLAS2013,labATLASNew,labCMSNew}.}
\footnotesize
\begin{tabular*}{80mm}{c@{\extracolsep{\fill}}ccc}
\toprule Experiment & \hphantom{0}Final state & \hphantom{0}($\sqrt{s}$/TeV, L/$fb^{-1}$) & \hphantom{0}$\mu$ \\
\hline
CMS\hphantom{0} & \hphantom{0}$\gamma\gamma$ & \hphantom{0}(7, 5.1)+(8, 19.6) & \hphantom{0}$0.77\pm0.27$ \\
CMS\hphantom{0} & \hphantom{0}ZZ & \hphantom{0}(7, 5.1)+(8, 19.6) & \hphantom{0}$0.92\pm0.28$ \\
CMS\hphantom{0} & \hphantom{0}WW & \hphantom{0}(7, 4.9)+(8, 19.5)\hphantom{0} & \hphantom{0}$0.68\pm0.20$\hphantom{0} \\
CMS\hphantom{0} & \hphantom{0}bb & \hphantom{0}(7, 4.9)+(8, 12.1)\hphantom{0} & \hphantom{0}$1.15\pm0.62$ \\
CMS\hphantom{0} & \hphantom{0}$\tau\tau$ & \hphantom{0}(7, 4.9)+(8, 19.4) & \hphantom{0}$1.10\pm0.41$ \\
\hline
ATLAS\hphantom{0} & \hphantom{0}$\gamma\gamma$ & \hphantom{0}(7, 4.8)+(8, 20.7) & \hphantom{0}$1.6\pm0.30$ \\
ATLAS\hphantom{0} & \hphantom{0}ZZ & \hphantom{0}(7, 4.6)+(8, 20.7) & \hphantom{0}$1.5\pm0.40$ \\
ATLAS\hphantom{0} & \hphantom{0}WW & \hphantom{0}(8, 13)\hphantom{0} & \hphantom{0}$1.4\pm0.60$\hphantom{0} \\
ATLAS\hphantom{0} & \hphantom{0}bb & \hphantom{0}(7, 4.7)+(8, 13)\hphantom{0} & \hphantom{0}$-0.4\pm1.00$ \\
ATLAS\hphantom{0} & \hphantom{0}$\tau\tau$ & \hphantom{0}(7, 4.6)+(8, 13) & \hphantom{0}$0.8\pm0.70$ \\
\bottomrule
\end{tabular*}
\vspace{0mm}
\end{center}


\clearpage


\begin{thebibliography}{90}

\vspace{3mm}
\bibitem{labFANHiggsF} F. Englert and R. Brout, Phys. Rev. Lett. {\bf13}, 321 (1964)
\bibitem{labFANHiggsH} P. W. Higgs, Phys. Rev. Lett. {\bf13}, 508 (1964)
\bibitem{labFANHiggsG} G. Guralnik, C. Hagen and T. Kibble, Phys. Rev. Lett. {\bf13}, 585 (1964)
\bibitem{labCMSHiggs} CMS Collaboration, Phys. Lett. B {\bf 716}, 30 (2012)
\bibitem{labATLASHiggs} ATLAS Collaboration, Phys. Lett. B {\bf 716}, 1 (2012)
\bibitem{labATLASNew} ATLAS Collaboration, ATLAS-CONF-{\bf2013-014},
\url{https://twiki.cern.ch/twiki/bin/view/AtlasPublic/HiggsPublicResults}
\bibitem{labCMSNew} CMS Collaboration, CMS-HIG-{\bf13-005},
\url{https://twiki.cern.ch/twiki/bin/view/CMSPublic/PhysicsResultsHIG}
\bibitem{labTevatronHiggs} CDF and D0 Collaborations, Submitted to Phys. Rev. D, 25 Mar 2013, arXiv:1303.6346v1[hep-ex]
\bibitem{labFANSUSYP}  P. Fayet, Phys. Lett. B {\bf64}, 159 (1976)
\bibitem{labFANSUSYF}  P. Fayet, Phys. Lett. B {\bf69}, 489 (1977).
\bibitem{labFANSUSYR}  G. R. Farrar and P. Fayet, Phys. Lett. B {\bf76}, 575 (1978)
\bibitem{labFANSUSYM}  G. F. Giudice, M. A. Luty, H. Murayama and R. Rattazzi, J. High Energy Phys. {\bf12}, 027 (1998)
\bibitem{labFANMSSM1}  H. P. Nilles, Phys. Rept. {\bf110}, 1 (1984)
\bibitem{labFANMSSM2}  H. E. Haber and G. L. Kane, Phys. Rept. {\bf117}, 75 (1985)
\bibitem{labFANMSSM3}  R. Barbieri, Riv. Nuovo Cim. {\bf11}, 1 (1988)
\bibitem{labmuproblem} G. F. Giudice and A. Masiero, Phys. Lett. B {\bf 206}, 480 (1998)
\bibitem{labmuproblemFAN}  J. E. Kim and H. P. Nilles, Phys. Lett. B {\bf138}, 150 (1984).
\bibitem{labNMSSM} U. Ellwanger, C. Hugonie and A. M. Teixeira, Phys. Rept. {\bf 496} (2010):1-77
\bibitem{labLEPHiggs} R. Barate et al, Phys. Lett. B {\bf 565}, 61 (2003)
\bibitem{labnewUEll} G. B\'elanger, U. Ellwanger, J. F. Gunion, Y. Jiang, S. Kraml, J. H. Schwarz, JHEP {\bf 1301} (2013) 069
\bibitem{labNSMMS1} J. Cao, Z. Heng, D. Li and J. M. Yang, Phys. Lett. B {\bf 710}, 665 (2012)
\bibitem{labNSMMS2} J. F. Gunion, Y. Jiang and S. Kraml, Phys. Lett. B {\bf 710}, 454 (2012)
\bibitem{labNSMMS3} S. F. King, M. Muhlleitner and R. Nevzorov, Nucl. Phys. B {\bf 860}, 207 (2012)
\bibitem{labMarcin} M. Badziak, M. Olechowski and S. Pokorski, J. High Energy Phys. {\bf1306}, 043 (2013)
\bibitem{labAndrea} R. Barbieri, D. Buttazzo, K. Kannike, F. Sala and A. Tesi, Phys. Rev. D {\bf87}, 115018 (2013)
\bibitem{labFANLHCSUSY1} Paul de Jong for ATLAS and CMS Collaboration, Proceedings of Physics in Collision 2012,
ATL-PHYS-PROC-{\bf2012-244}, [arXiv:1211.3887 [hep-ex]]
\bibitem{labFANLHCSUSY2} CMS Collaboration, CMS-SUS-{\bf13-011}, CMS-SUS-{\bf13-003}, PAS-SUS-{\bf13-017}
\bibitem{labFANLHCSUSY3} ATLAS Collaboration, ATLAS-CONF-{\bf2013-093}, ATLAS-CONF-{\bf2013-062}, ATLAS-CONF-{\bf2013-089}
\bibitem{labPLANK} Planck Collaboration, Submitted to Astronomy and Astrophysics, 20 Mar 2013, [arXiv:1303.5076 [astro-ph.CO]]
\bibitem{labDM} D. Stockinger, J. Phys. G {\bf 34}, R45 (2007)
\bibitem{labNMSSMTools} \url{http://www.th.u-psud.fr/NMHDECAY/nmssmtools.html}
\bibitem{labNMSSMToolsa} U. Ellwanger, J. F. Gunion and C. Hugonie, JHEP {\bf 0502} (2005) 006
\bibitem{labNMSSMToolsb} U. Ellwanger and C. Hugonie, Comput. Phys. Commun. {\bf 175} (2006) 290
\bibitem{labNMSSMToolsc} G. Belanger, F. Boudjema, C. Hugonie, A. Pukhov and A. Semenov, JCAP {\bf 0509} (2005) 001
\bibitem{labNMSSMToolsd} U. Ellwanger and C. Hugonie, Comput.Phys.Commun. {\bf177} (2007) 399-407
\bibitem{labNMSSMToolse} U. Ellwanger, C.-C. Jean-Louis and A.M. Teixeira, JHEP  {\bf 0805} (2008) 044
\bibitem{labNMSSMToolsf} Debottam Das, Ulrich Ellwanger, Ana M. Teixeira, arXiv:1106.5633 [hep-ph].
\bibitem{labNMSSMToolsg} M. Muhlleitner,  A. Djouadi, Y. Mambrini, Comput.Phys.Commun. {\bf 168} (2005) 46-70
\bibitem{labMCMC} B. Dumont, J. F. Gunion and S. Kraml, Phys. Rev. D {\bf 89}, 055018 (2014)
\bibitem{labHBound} P. Bechtle  et al., Comput. Phys. Commun. {\bf181}
(2010):138-167; P. Bechtle  et al., Comput. Phys. Commun. {\bf182}
(2011):2605-2631; P. Bechtle  et al., PoS CHARGED{\bf2012} (2012)
024; P. Bechtle  et al., BONN-TH-{\bf2013-21}, DESY {\bf13-110},
arXiv:1311.0055[hep-ph]
\bibitem{labHSignal} P. Bechtle, S. Heinemeyer, BONN-TH-{\bf 2013-07}, DESY {\bf 13-078}, 3
(2013), arXiv:1305.1933 [hep-ph]; O. Stal, T. Stefaniak,
BONN-TH-{\bf2013-20}, arXiv:1310.4039 [hep-ph]
\bibitem{lanNMSSMLHCFAN}  G. B\'elanger, B. Dumont, U. Ellwanger, J. F. Gunion, S. Kraml, Phys.Rev. {\bf D88} (2013) 075008
\bibitem{labSLHA} B. Allanach et al. Comput. Phys. Commun. {\bf 180}, 8 (2009)
\bibitem{SchmidtHoberg} K. Schmidt-Hoberg, F. Staub, JHEP {\bf 1210} (2012) 195
\bibitem{Choi} K. Choi, S. H. Im, K. S. Jeong, M. Yamaguchi, JHEP {\bf 1302} (2013) 090
\bibitem{labCoup} S. F. King, M. Muhlleitner, R. Nevzorov3, K. Walz, Nuclear Physics B  {\bf 870} 2 (2012)
\bibitem{Ellwanger2010nf} U. Ellwanger, Phys. Lett. B {\bf 698} (2011) 293
\bibitem{labBPHY1} A. J. Buras, P. H. Chankowski, J. Rosiek, L. Slawianowska, Nucl. Phys. B {\bf 659}, 3(2003)
\bibitem{labBPHY2} M. Misiak et al. Phys. Rev. Lett. {\bf 98}, 022002 (2007)
\bibitem{labBPHY3} G. Hiller, Phys. Rev. D 70 (2004) 034018 [arXiv:hep-ph/0404220]. 
\bibitem{labBPHY4} F. Domingo and U. Ellwanger, JHEP {\bf 0712} (2007) 090 [arXiv:0710.3714 [hep-ph]]. 
\bibitem{labFANRA} T. Elliott, S. F. King, P. L. White, Phys. Rev. D {\bf49} , 12 (1994):2435-2456
\bibitem{labFANWMAP} D. N. Spergel, R. Bean, Astrophys. J. Suppl. {\bf170}, 5 (2007):377
\bibitem{labFANBR} LHC Higgs Cross Section Working Group,
\url{https://twiki.cern.ch/twiki/bin/view/LHCPhysics/CrossSections}
\bibitem{labCMS2013} CMS Collaboration, CMS-HIG-{\bf13-001},
\url{https://twiki.cern.ch/twiki/bin/view/CMSPublic/PhysicsResultsHIG}
\bibitem{labATLAS2013} ATLAS Collaboration, ATLAS-CONF-{\bf2013-007},
\url{https://twiki.cern.ch/twiki/bin/view/AtlasPublic/HiggsPublicResults}
\bibitem{labDoubleHiggs} CMS Collaboration, CMS-HIG-{\bf13-016},
\url{https://twiki.cern.ch/twiki/bin/view/CMSPublic/PhysicsResultsHIG}
\end{thebibliography}
\end{document}